\documentclass{aa}  
\usepackage{graphicx}
\usepackage{txfonts}
\usepackage{float}
\usepackage{xcolor}
\usepackage{natbib,twoopt}
\usepackage[breaklinks=true]{hyperref} 
\bibpunct{(}{)}{;}{a}{}{,} 
\makeatletter
\newcommandtwoopt{\citeads}[3][][]{\href{http://adsabs.harvard.edu/abs/#3}%
{\def\hyper@linkstart##1##2{}%
\let\hyper@linkend\@empty\citealp[#1][#2]{#3}}}
\newcommandtwoopt{\citepads}[3][][]{\href{http://adsabs.harvard.edu/abs/#3}%
{\def\hyper@linkstart##1##2{}%
\let\hyper@linkend\@empty\citep[#1][#2]{#3}}}
\newcommandtwoopt{\citetads}[3][][]{\href{http://adsabs.harvard.edu/abs/#3}%
{\def\hyper@linkstart##1##2{}%
\let\hyper@linkend\@empty\citet[#1][#2]{#3}}}
\newcommandtwoopt{\citeyearads}[3][][]%
{\href{http://adsabs.harvard.edu/abs/#3}
{\def\hyper@linkstart##1##2{}%
\let\hyper@linkend\@empty\citeyear[#1][#2]{#3}}}
\makeatother

\usepackage[normalem]{ulem}
\usepackage{soul}

\begin{document}

   \title{The effect of heavy ions on the dispersion properties of kinetic Alfvén waves in astrophysical plasmas}


   \author{N. Villarroel-Sep\'{u}lveda
          \inst{1}
          \and
          R. A. L\'{o}pez \inst{2} 
          \and 
          P. S. Moya \inst{1}
          }

   \institute{Departmento de F\'{\i}sica, Facultad de Ciencias,
Universidad de Chile, Las Palmeras 3425, 7800003, Ñuñoa, Santiago, Chile\\
              \email{nicolas.fvsv@gmail.com, pablo.moya@uchile.cl}\\
         \and
            Departamento de F\'{\i}sica, Universidad de Santiago de Chile, Usach, 9170124, Santiago, Chile\\
             \email{rodrigo.lopez.h@usach.cl}
            }

 \abstract
   {Spacecraft measurements have shown Kinetic Alfvén Waves propagating in the terrestrial magnetosphere at lower wave-normal angles than predicted by linear Vlasov theory of electron-proton plasmas. To explain these observations, it has been suggested that the abundant heavy ion populations in this region may have strong, non-trivial effects that allow Alfvénic waves to acquire right-handed polarization at lower angles with respect to the background magnetic field, as in the case of typical electron-proton plasma.}
   {We study the dispersion properties of Alfvénic waves in plasmas with stationary phase-space distribution functions with different heavy ion populations. Our extensive numerical analysis has allowed us to quantify the role of the heavy ion components on the transition from the left-hand polarized electromagnetic ion-cyclotron (EMIC) mode to the right-hand polarized kinetic Alfvén wave (KAW) mode.  }
   {We used linear Vlasov-Maxwell theory to obtain the dispersion relation for oblique electromagnetic waves. The dispersion relation of Alfvén waves was obtained numerically by considering four different oxygen ion concentrations ranging between $0.0$ and $0.2$ for all propagation angles,  as a function of both the wavenumber and the plasma beta parameter. }
   {The inclusion of the heavy O${}^{+}$ ions is found to considerably reduce the transition angle from EMIC to KAW both as a function of the wave number and plasma beta. With increasing O${}^{+}$ concentrations, waves become more damped in specific wavenumber regions. However, the inclusion of oxygen ions may allow weakly damped KAW to effectively propagate at smaller wave-normal angles than in the electron-proton case, as suggested by observations.} 
   {}
   \keywords{ plasma --
                waves --
                polarization
               }

  \maketitle
%
\section{Introduction}

Kinetic Alfvén waves (KAWs) are Alfvénic waves that propagate at large wave-normal angles, when the mode turns compressive and develops a large parallel electric field due to kinetic effects, which give this mode its name. These characteristic kinetic effects become relevant under two conditions in typical electron-proton plasma: 1) in the hot electron case ($\beta>m_e/m_p$) as the perpendicular wavelength reaches the order of the scale of the protons’ gyroradius \citep{Hasegawa1977,gary_1986,Hollweg1999} and 2) in the cold electron case ($\beta<m_e/m_p$) as it reaches the electron's inertial length \citep{goertz_boswell_1979, lysak_lotko_1996, tamrakar_varma_tiwari_2019, lysak_2023}. In this latter case, the waves can also be referred to as inertial Alfvén waves \citep{assis_silva_carvalho_2020}. Kinetic Alfvén waves are dispersive in the subproton scale, meaning they can effectively interact with the plasma particles in this domain \citep{gary_1986, Yang2005,Nandal2016}. The KAW is particularly relevant for the study of space and astrophysical plasmas, as many authors have suggested that it plays a crucial role in several kinetic processes, such as the energy transfer from larger scales toward smaller electron scales through a turbulence cascade \citep{Sahraoui2012, Salem_2012, Boldyrev2012, Boldyrev_2013,  Nandal2016}, magnetic reconnection \citep{Dai2017,Boldyrev2019, Cheng2020}, and plasma particle energization in the magnetosphere \citep{Artemyev2015, Tian2022} and solar atmosphere \citep{wu_yang_2006,artemyev_zimovets_rankin_2016}.

The large-amplitude parallel electric perturbations enable KAWs to accelerate charged particles along geomagnetic field lines, which allows for the strong acceleration and subsequent energization of electrons, in particular \citep{lysak_lotko_1996, Tian2022, Zhang2022, lysak_2023}. Among other observations, satellite measurements have provided direct evidence of electron acceleration by KAWs in the plasma sheet boundary layer \citep{Wygant2002, chaston_bonnell_2012, Zhang2022} and the equatorial inner magnetosphere \citep{angelopoulos_2002, Tian2022}. The spectral properties of KAW also provide mechanisms for the anomalous transport \citep{hasegawa_mima_1978, izutsu_hasegawa_nakamura_fujimoto_2012} and heating of ions \citep{chen_lin_white_2001}, all phenomena that have also been observed in the inner magnetosphere associated with magnetic shock impacts \citep{Malaspina2015}, substorms~\citep{Chaston2005}, and geomagnetic storms~\citep{Moya2015}.  With respect to this central role in regulating large-scale to electron-scale physical processes in space plasmas, the characterization of KAWs and their properties becomes a task of great relevance for our understanding of magnetospheric plasma and the role of wave-particle interactions in plasma phenomena.

For small propagation angles with respect to the mean magnetic field, the Alfvénic mode in the same frequency range as KAWs corresponds to the non-compressive electromagnetic ion-cyclotron (EMIC) mode. As the name suggests, the EMIC mode allows strong cyclotron resonance with the ions in the plasma, but not with the electrons \citep{Gary2004}. This feature is a consequence of the left-handed polarization of EMIC waves in the plasma frame, although the polarization of these waves can also be linear as a limit to the left-hand elliptical polarization \citep{gary_1986}. However, when the wave vector develops a large component perpendicular to the background magnetic field, these waves become compressive as they acquire strong electric field fluctuations parallel to the mean magnetic field, as well as shifting from left-hand to right-handed polarization (or linear, this time as a limit to the right-hand elliptical polarization) in the plasma frame \citep{gary_1986,Gary2004,Hollweg1999, hunana__2013, Nandal2016} as the plasma's diamagnetic current becomes larger than its Hall current \cite{Narita2020}. So, for fixed values of the wavenumber, plasma beta, and other parameters, the left-hand polarized EMIC mode transitions to the right-handed KAW mode as the propagation angle increases. As shown by \citet{gary_1986}, this transition angle depends on the wavenumber, and the plasma beta parameter (the ratio between thermal and magnetic pressure in the system). These waves have become the object of study in a wide range of plasma environments, with a particular interest in the heliospheric and magnetospheric environments, since spacecraft observations have consistently shown that the aforementioned Alfvénic modes propagate in the solar wind and different regions of Earth's magnetosphere, such as the magnetopause, plasma sheet, magnetosheath, and the inner magnetosphere \citep{Wygant2002,chaston_bonnell_2012,Chaston2014,Chaston2015,Perschke2013,Malaspina2015,saikin_2015,ni_cao_shprits_summers_gu_fu_lou_2018, khotyaintsev_2021,Tian2022,Zhang2022}. Furthermore, the excitation and propagation of KAW and other Alfvénic waves have been theoretically proposed in different space plasma environments, such as planetary magnetospheres resembling  those of Mars and Saturn \citep{singh_2019,barik_singh_lakhina_2021}; dusty plasmas, as in planetary disks and cometary tails \citep{jatenco-pereira_2014,saini_singh_bains_2015}; neutron stars or black hole surrounding plasmas \citep{nattila_2022}; and many other astrophysical plasma environments \citep{sah_2010}. The excitation of KAW and ion-acoustic waves by hot ion beams and velocity shear, through both resonant and non-resonant instabilities, has also been extensively and in depth in plasmas composed of hot electrons and cold ions \citep{lakhina_1990, lakhina_2008, barik_singh_lakhina_2019a, barik_singh_lakhina_2019b,  barik_singh_lakhina_2019c, barik_singh_lakhina_2020}. This mechanism for the excitation of ULF waves has been linked to the study of the polar cusp region of Earth's magnetosphere, but the results are more general and can be applied to any plasma of these characteristics.

Most studies on KAWs have focused on the typical case of ideal electron-proton plasma (see, for example, \citet{gary_1986, Hollweg1999, Salem_2012}). Space plasmas are, however, constituted by different plasma species, not only electrons and protons. Since the dispersion properties for oblique waves in a collisionless plasma depend on the different ion relative density \citep{gary_1986,Stix1992,Vinas_2000,Lopez_2017,Chen2021,MOYA2021}, these populations often result significant enough not to be neglected. For the particular case of the inner magnetosphere, spacecraft observations have shown KAWs propagating at lower angles with respect to the background magnetic field than predicted by the electron-proton theory applied to this environment. Previous studies have proposed that this puzzling behavior may be explained by the role played by the presence of heavy ions in this region of space \citep{Moya2015,MOYA2021,Moya2022}, where the concentration of He$ ^+$ ions range approximately from 5\% up to 20\% and of O$^+$ ions from 20\% up to 80\% of the total ion population depending on the level of geomagnetic activity~\citep{Jahn2017}, with higher O${}^+$ ion concentrations tightly linked to strong geomagnetic activity \cite{yue_2019}. According to \citet{MOYA2021,Moya2022}, the heavy ions introduce non-trivial changes to the susceptibility of the plasma, subsequently allowing for the presence of KAWs at lower angles than those predicted using an electron-proton plasma approximation.

The relevance of heavy ions in the different physical processes that occur in space and astrophysical plasmas is not restricted to the case of the terrestrial magnetosphere. Populations of heavy ions such as oxygen 
and nitrogen 
are present, in varying concentrations, in the atmospheres of Venus \citep{persson_2019}, Mars \citep{leblanc__2017}, Saturn \citep{hartle_2006, singh_2019} and its moon Titan \citep{cravens_2008, gronoff_2009}, and Jupiter \citep{roussos_2021} and its moon Juno, where sulfur ions are also an important component \citep{kim_2020}, and are likely components of the magnetospheres of close-in exoplanets \citep{johansson_mueller_motschmann_2010}.  The solar wind and other regions of the heliosphere are characterized by relatively high abundances of alpha particles, as well as smaller populations of heavier ions such as oxygen, nitrogen, neon, sulfur, silicon, and heavy metals \citep{bame_asbridge_feldman_montgomery_kearney_1975,young_2005a, cohen_mason_2017, zangrilli_giordano_2020, mason_2021}. Oxygen, nitrogen, 
 and heavy metal ions 
have also been found to be present in the accretion disks of black holes and active galactic nuclei \citep{fields_mathur_krongold_williams_nicastro_2007,Deprince_2019,Deprince2_2020}, while all of the previously mentioned atoms have also been observed in planetary nebulae at high abundances and various stages of ionization, along with carbon, sulfur, silicon, and other ion species \citep{pottasch_2007}. Some of the solar wind ions, such as oxygen and sulfur, are also important components of cometary plasmas in conjunction with molecular ions, which are formed through the ionization of neutral gas \citep{balsiger_1986}. Plasma simulation studies that consider some of the previously mentioned environments show that the presence of heavy ions effectively affects the plasma dynamics and excitation of electromagnetic fluctuations \citep{gary_madland_omidi_winske_1988, johansson_mueller_motschmann_2010,Moya2014}.

This relevance of heavy ions on the properties of electromagnetic waves in plasmas lays the groundwork for an in-depth analysis of the effect of these ions on the transition from left-hand  polarized to right-hand polarized Alfvén waves in a collisionless plasma as the propagation angle progresses from parallel to perpendicular. Since the cyclotron motion of charged particles in a background magnetic field depends only on the strength of the field and the charge-to-mass ratio of the particles, the latter quantity acquires particular importance in the properties of oblique waves, determining the resonance frequencies, among other aspects of the plasma wave dynamics \citep{wu_yang_2006,kumar_2017}. Although different astrophysical environments are intrinsically dissimilar when it comes to their composition, it is interesting to note that many of the most abundant heavy ions present in astrophysical plasmas (see references in the previous paragraph for deeper insights) have similar charge-to-mass ratios, N${}^+$  \citep{pottasch_2007, cravens_2008, gronoff_2009, singh_2019}, O${}^+$  \citep{balsiger_1986, hartle_2006, pottasch_2007,cravens_2008, leblanc__2017, Deprince_2019, persson_2019, singh_2019, kim_2020, roussos_2021}, S${}^{2+}$ \citep{pottasch_2007, kim_2020, roussos_2021}, Cl${}^{2+}$ \citep{pottasch_2007},  and even Fe${}^{3+}$ \citep{Deprince2_2020} and Fe${}^{4+}$ \citep{Deprince2_2020} have charge/mass ratios between $r_H/19$ and $r_H/14$, where $r_H=e/m_H$ is the charge/mass ratio of a H${}^+$ ion. Thus, the plasma wave dynamics of a multi-ion plasma composed of H${}^+$ and any of the aforementioned heavy ions may be somewhat similar.

In this manuscript, we aim to elucidate how the inclusion of heavy ions affects the transition angle from the left-hand polarized electromagnetic ion-cyclotron (EMIC) waves to the right-hand polarized KAW. As O${}^+$ ions happen to be one of the most common ion species in many of the astrophysical plasmas mentioned above and because it is the most abundant heavy ion species in Earth's inner magnetosphere, where their concentration ranges from 20\% to 50\% of the total ion population \citep{Jahn2017}, we consider the case study of an e${}^-$-H${}^+$-O${}^+$ plasma with magnetospheric parameters. Our goal is to examine the potential role of heavy ions on the dispersion properties of Alfvénic waves in the kinetic regime in a collisionless magnetized plasma. We  use linear Vlasov-Maxwell theory considering different concentrations of oxygen ions to analyze the dispersion relation and polarization extensively, as a function of the wave number and plasma beta parameter, in hopes of providing a theoretical explanation for the observation of KAWs propagating at oblique angles in the terrestrial magnetosphere. We use this information  to make predictions about KAW propagation in other astrophysical plasma environments where heavy ions are ubiquitous.


\section{Plasma model for a multi-ion collisionless plasma}


In this study, we consider a collisionless plasma composed of electrons, H${}^+$ ions, and a smaller (but significant) population of the heavier O${}^+$ ions, whose charge/mass ratio is $\sim r_H/16$. We impose the quasi-neutrality condition $\sum_{s} q_s n_s=0$, where $q_s$ and $n_s$ represent the charge and density of the species ``$s$.'' Since all species that compose the plasma have elementary charge $\pm e$, this condition is fulfilled when $n_0=n_{H^+}+n_{O^+}=n_e$, with $n_0$ and $n_e$ as the total ion and electron densities, respectively. We can write this relation in terms of the ion concentrations, such that $n_{H^+} = n_0\,\eta_{H} $, $n_{O^+}=n_0\,\eta_{O^+}$, and $\eta_{H} + \eta_{O^+}=1$. Here, $\eta$ represents the relative density of each species. In the following, we analyze the properties of Alfvénic waves for concentrations of oxygen ions given by $\eta_{O^+}=0.05$, $\eta_{O^+}=0.10$ and $\eta_{O^+}=0.20$ and compare them to the typical electron-proton case. A plasma of these characteristics can be considered a simplified model for astrophysical plasmas where the phase-space distribution function of the particles is stationary, and the heavy ion populations are predominantly composed of only one species. This is the case, for example, for certain regions of Earth's inner magnetosphere \citep{Jahn2017}, Saturn's inner plasmasphere \citep{young_2005b}, or the plasma torus in the jovian magnetosphere, where O${}^+$ and S${}^{++}$ (which have roughly the same charge-to-mass ratio) make up most of the heavy ion population \citep{yoshioka_2017}. This is also considering that the three of these planets have dynamo-generated magnetic fields with similar dipole moments \citep{russell_dougherty_2010}, making them comparable plasma environments.

We consider isotropic temperatures, such that $T_{\parallel s}=T_{\perp s}=T_s$, for all species to reduce the space of parameters to only the wave number, plasma beta, and oxygen ion concentration. These three quantities are the main focus of our current investigation (for an analysis of the dependence of the dispersive properties of Alfvénic waves on the temperature anisotropy of the species in a similar context, see \citet{MOYA2021}).

Concerning the plasma beta parameter, observations over the full range of magnetic local time in the ring current region of the inner magnetosphere have shown that its value for protons usually ranges from $\beta_{H^+}\sim 10^{-3}$ to $\beta_{H^+}\sim 2$, both during quiet and active geomagnetic times. While some observations with $\beta_{H^+}$ values greater than $4$ have also been recorded in this region, these events correspond to only about 0.2\% of the total of high beta observations, according to \citet{Cohen2017}. In contrast, the probability of a high beta event with $\beta_{H^+}\leq3$ rounds to 98\%. Thus, to study the dependence of the dispersion properties on the parallel plasma beta parameter, we limited our domain of study to $10^{-3}\leq\bar{\beta}_{ H^{+}}\leq 4$, where $\bar{\beta}_{ s}=8\pi n_0 k_B T_{s}/B_0^2$, with $B_0$ the background magnetic field, $T_{ s}$ is the temperature of species ``$s$'' in the direction parallel to the background magnetic field, and $k_B$ is the Boltzmann constant. This modified plasma beta parameter, $\bar{\beta}$, introduced by \citet{gary_1993}, accounts only for the parallel temperature of the species when the particle density stays fixed and is related to the regular beta parameter by $\beta_{s}=\eta_s\bar{\beta}_{ s}$. The plasma beta parameters of H${}^+$ ions observed in Saturn's magnetosphere are fully contained in this range, with $0<\beta_{H^{+}}<2$ \citep{thomsen_2010}.

Unlike other astrophysical environments such as the solar wind, particle populations within inner planetary magnetospheres usually do not drift significantly with respect to one another in the direction of the field lines. We therefore suppressed this drift for the sake of simplicity. By taking this aspect into consideration, as well as the temperature isotropy mentioned above, we can solve the dispersion relation for oblique waves by using the linear Vlasov-Maxwell theory for nondrifting Maxwellian velocity distributions given by:
\begin{equation}\label{Eq1}
    f_{s}(v_\parallel,v_\perp)=\frac{n_{s}\pi^{-3/2}}{\alpha_{s}^3 }\exp \Bigg\{{-\frac{v_\perp^2}{\alpha_{ s}^2}-\frac{v_\parallel^2}{\alpha_{ s}^2}}\Bigg\},
\end{equation}
for each particle species, where $\alpha_s=\sqrt{2k_BT_s/m_s}$ is the thermal speed of each species, with $m_s$ as the mass of each species. We consider waves propagating obliquely to the mean magnetic field $\mathbf{B}=B_0\hat{\mathbf{z}}$, with $\mathbf{k}=k_\perp\hat{\mathbf{x}}+k_\parallel\hat{\mathbf{z}} = k\sin\theta\hat{\mathbf{x}}+k\cos\theta\hat{\mathbf{z}}$, where $\theta$ is the propagation or wave-normal angle. From this, we obtain the dispersion relation:

\begin{align}
 \mathbf{D}_k\cdot\delta\mathbf{E}_k=0, \label{disp}
\end{align}
where $\mathbf{D}_k$ is the full dispersion tensor for oblique waves and $\delta\mathbf{E}_k$ represent the electric field eigenmodes (see Appendix A for further detail). With the information of the dispersion tensor, we can compute the polarization of the transverse component of the waves as taken with respect to the background magnetic field, as defined by \citet{Stix1992}, using:


\begin{align}
   P(k)=i\frac{\delta E_{kx}}{\delta E_{ky}}\label{polStix}
.\end{align} 
As usual in the context of plasma physics, right-handed polarization is defined by a timewise gyration of the fields in the direction of the background magnetic field according to the right-hand rule, whereas in left-hand polarized waves, the fields gyrate in the opposite sense. Thus, right-hand (left-hand) polarized waves have fields gyrating in the sense of the Larmor gyration of negatively (positively) charged particles.\\
Considering bounded electric perturbations, a linear polarization is achieved whenever $P(k)=0$ or $P(k)=\pm\infty$. Circular polarization occurs when $P(k)=\pm1$; where the plus sign implies right-handed polarization and the minus sign is associated with left-handed polarization. Other values of finite $\textnormal{Re}(P)>0$ 
and $\textnormal{Re}(P)<0$ indicate a right-handed and left-handed elliptical polarization, respectively. 
It is crucial to note, however, that the main interest of this study is the sign of polarization and not its exact value. Because of this, polarization values in the following results are shown normalized to their highest values, which sometimes correspond to numerical infinity, in Figures \ref{fig:disp_angles} and \ref{fig:disp_angles_beta}. To avoid dividing by very big numbers while normalizing the polarization in Figures \ref{fig:heat_k} and \ref{fig:heat_beta}, which would result in the maps showing polarization values very close to 0 everywhere, except for very hot (or cold) small pockets, we chose to set the most extreme values of the polarization at $\text{Re}(P)=\pm100$ and normalize to this number. Therefore, $\text{Re}(P)=\pm1$ will not imply circular polarization as for the results shown and discussed hereafter.

We solved the dispersion relation of Alfvénic waves using the DIS-K solver \footnote{The full code is publicly available and can be found at \url{https://github.com/ralopezh/dis-k}.} \citep{Lopez_2021}, in the limit of isotropic Maxwellian distributions as given by Eq.~(\ref{Eq1}). We identified the Alfvén waves and differentiated them from other solutions that possess right-handed polarization at near-perpendicular angles by analyzing the phase velocity of the waves in the MHD limit (see Appendix B). For a fixed value of the beta parameter, we obtained the real and imaginary parts of the wave's frequency normalized to the protons' cyclotron frequency $\Omega_{p}$, as well as their polarization -- all as a function of the wavenumber normalized to the inverse of the proton inertial length $\omega_{pp}/c$, with $\omega_{pp}$ as the protons' plasma frequency and $c$ as the speed of light in vacuum. As a validation task, this routine was able to accurately reproduce Fig.  3 of \citet{gary_1986} for the transition angle from left-hand polarized to right-hand polarized Alfvén waves at $kc/\Omega_p=0.05$ as a function of the beta parameter in an isotropic electron-proton plasma (see Fig.  \ref{fig:polk05} in Appendix C). 

For all calculations, we considered isothermy between the plasma species, such that $\bar{\beta}_e=\bar{\beta}_{H^+}=\bar{\beta}_{O^+}=\bar{\beta}$. Since $\eta_e=1$, it follows that $\beta_e=\bar{\beta}$. Thus,  the beta parameter for electrons $\beta_e$ is used in the following sections as a variable from which the beta parameter of each species can be obtained, given the different ion concentrations.

%

\section{Results}

\subsection{Effect of oxygen ions on the dispersion relation of Alfvénic waves}

\begin{figure}[ht]
    \centering
    \includegraphics[width=0.5\textwidth]{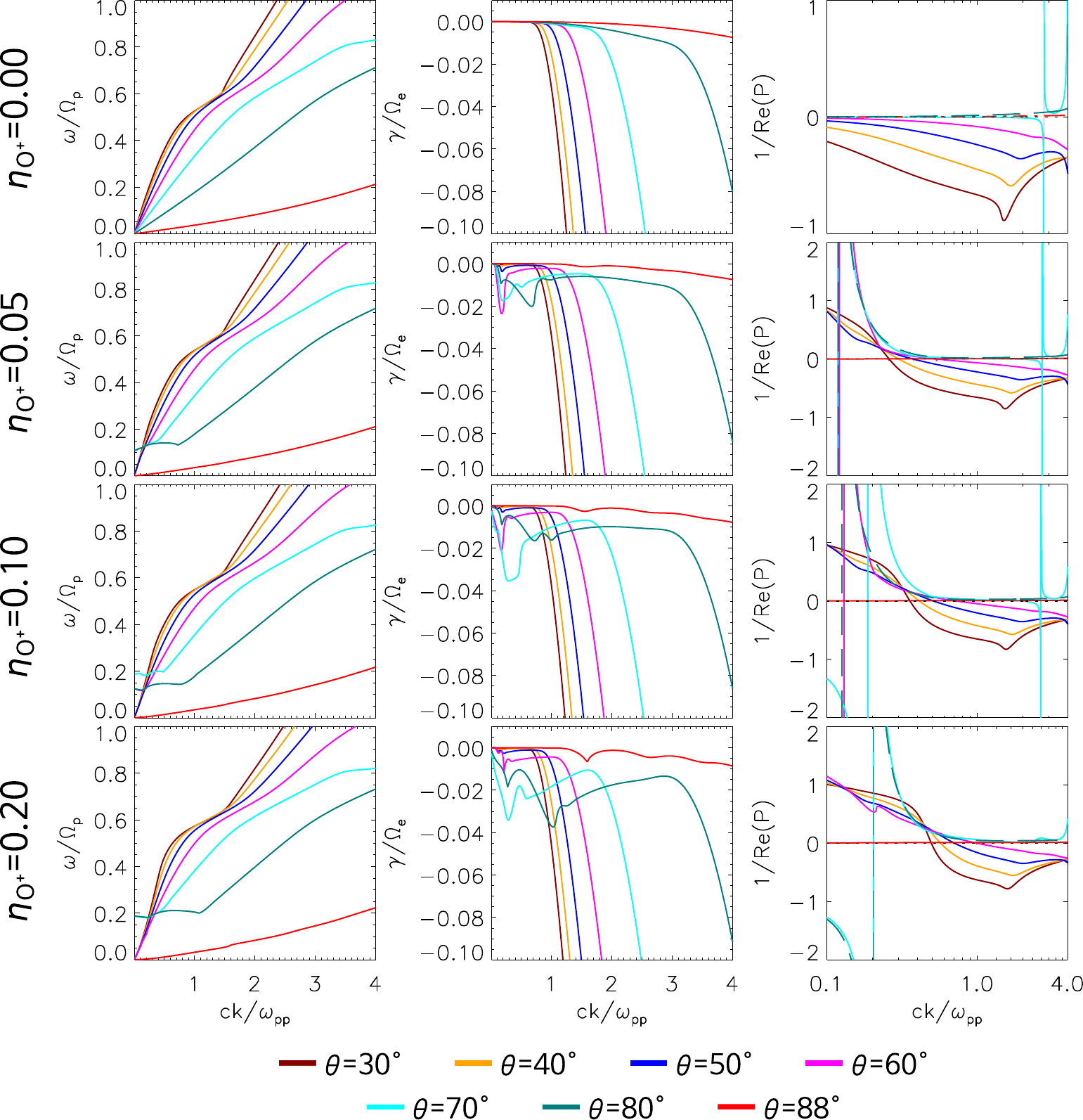}
    \caption{Real part (first column), imaginary part (second column) of the frequency, and inverse real part of the polarization (third column) as a function of the wavenumber for Aflvénic waves propagating at 30º, 40º, 50º, 60º, 70º, 80º, and 88º with respect to the background magnetic field, for oxygen ion concentrations of $\eta_{O^+}=0.00$ (first row), $\eta_{O^+}=0.05$ (second row), $\eta_{O^+}=0.10$ (third row), and $\eta_{O^+}=0.20$ (fourth row). For all curves, the plasma beta parameter is $\beta_e=0.1$. }
    \label{fig:disp_angles}
\end{figure}
\begin{figure}[ht]
    \centering
    \includegraphics[width=0.5\textwidth]{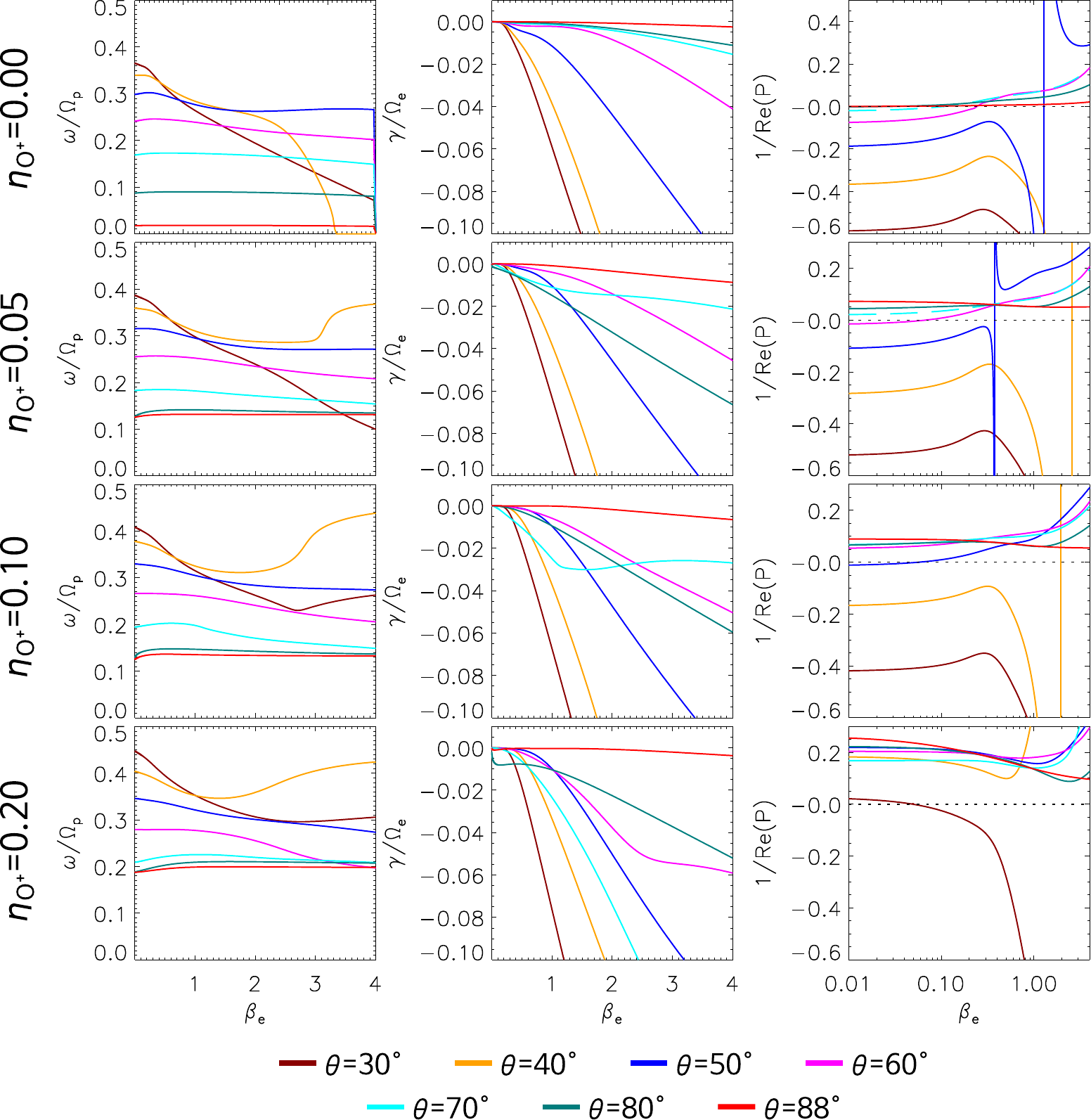}
    \caption{Real and imaginary parts of the frequency and inverse real part of the polarization for different oxygen ion concentrations and wave-normal angles as a function of the electron plasma beta parameter, displayed in the same format as Fig.  \ref{fig:disp_angles}. For all curves, the normalized wavenumber is set at $ck/\omega_{pp}=0.5$.}
    \label{fig:disp_angles_beta}
\end{figure}

Figure \ref{fig:disp_angles} shows the real (left) and imaginary (center) parts of the frequency, and the normalized polarization of Alfvénic waves versus the normalized wave number, for various oxygen ion concentrations and different propagation angles between 30$^{\circ}$ and 90$^{\circ}$. We did not include lower angles because (in those cases) the waves lose their positive polarization; this means that all solutions correspond to EMIC waves. In the figure, the top row shows the typical electron-proton case with zero oxygen concentration. The concentration increases for the subsequent rows, from $\eta_{O^+} = 0.05$ to $\eta_{O^+} = 0.2$. These results show a substantial modification of the dispersion properties of the waves with the inclusion of oxygen ions. One of the more apparent modifications is that, unlike in the electron-proton case, the real part of the frequency does not pass through zero for some wave-normal angles. This phenomenon, seen in the frequency curves for propagation angles of 70º and 80º for $\eta_{O^+}\geq0.05$, is a consequence of the separation of the solution into two frequency bands, one asymptotic to the oxygen's gyroradius (which is not plotted in the Figure \ref{fig:disp_angles}) and the other to the proton's gyroradius. We only included the solutions that address the proton's gyrofrequency, as this is the one we can feasibly compare to the solutions obtained in the electron-proton plasma, in line with the main objective of this study. This separation of the Alfvén mode into different frequency bands has been previously discussed by \citet{isenberg_1984,Gomberoff1991,Yang2005,Moya2015,Moya2022}, among others. In addition, it is a consequence of the apparition of forbidden frequency domains in the neighborhoods of the different ion's gyrofrequencies. We also observe a small wave number domain where the mode becomes slightly damped -- a phenomenon that is not present in the first case. This increased damping due to the presence of O${}^+$ (or other heavy ions) has been previously observed by \cite{tamrakar_varma_tiwari_2019} and associated with particle energization by the waves. Consistently with recent results \citep{MOYA2021}, Alfvénic waves in a plasma infused with heavy ions acquire positive polarization in a wave number domain that grows with the propagation angle. Moreover, it is important to note that this is not the case in an electron-proton plasma. Finally, from Figure \ref{fig:disp_angles}, we also observe that the wave-number domain in which the polarization is positive increases in extension as the oxygen ion concentration increases, with the wave propagating at 70$^\circ$ (purple curves), losing its negative polarization entirely in the given wave number domain for $\eta_{O^+}=0.2$. 

Figure \ref{fig:disp_angles_beta} displays the dispersion properties of Alfvénic waves, this time as a function of the plasma beta parameter for a fixed wavenumber of $ck/\omega_{pp}=0.5$. This value of the wavenumber (approximately $0.1c/\omega_{pp}$, where $c/\omega_{pp}$ is the proton's inertial length) will be considered a lower limit to the kinetic regime. This is supported by the fact that at oblique angles, the kinetic solutions depart from their MHD counterparts at approximately this wave number (see Figure \ref{fig:B1} for propagation at 50$^\circ$ and 60$^\circ$). As a result, kinetic effects should be observable at the chosen value for this quantity.


From Figure \ref{fig:disp_angles_beta}, we find that the real part of the frequency becomes consistently more significant with increasing oxygen ion concentration for larger propagation angles. We also observe that when no oxygen ions are present, the frequency curve for propagation at 30º tends to decrease consistently with the value of the plasma beta parameter, as does the curve for propagation at 40º until the frequency reaches 0. This behavior disappears as the concentration of oxygen ions becomes higher; we can no longer see this steady decrease in the real part of the frequency for propagation at 40º when $\eta_{O^+}=0.05$. The same holds true for propagation at 30º when $\eta_{O^+}$ reaches 0.1. Moreover, the most interesting aspect regarding the effect of oxygen ion populations on the spectral properties of the waves becomes evident when we analyze the third column of the figure. We note that most of the initially negatively polarized waves acquire positive polarization as the beta parameter increases, even for low oxygen ion concentrations. This result is consistent with previous works \citep{gary_1986}.

\subsection{The Transition from EMIC to Kinetic Alfvén Waves}

\begin{figure*}[ht]
    \centering
    \includegraphics[width=\textwidth]{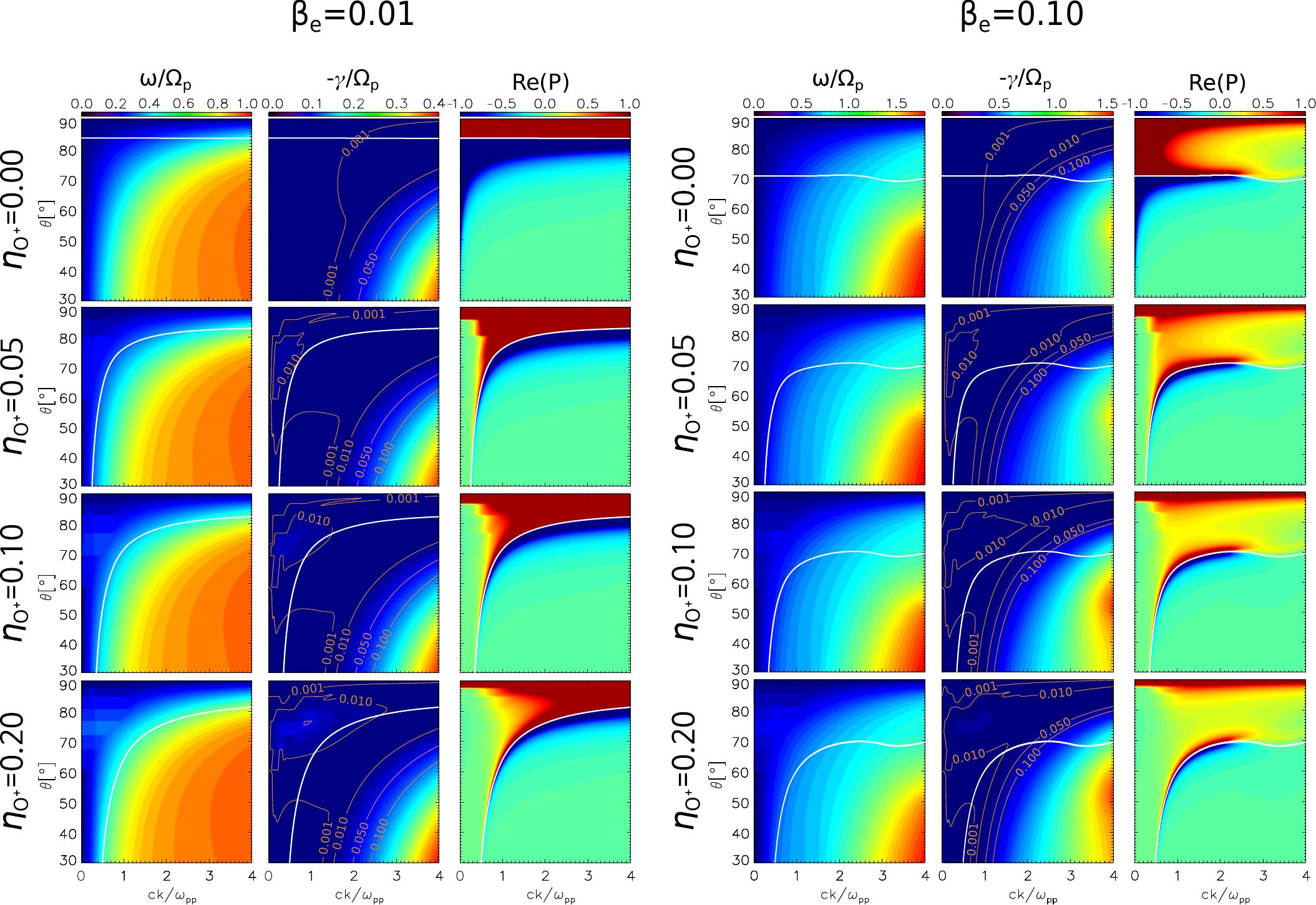}
    \caption{Heat maps of the real part of the normalized frequency $\omega/\Omega_p$, negative normalized damping rate $-\gamma/\Omega_p$, and real polarization $P$ for Alfvénic waves as functions of the propagation angle and wavenumber for the four oxygen ion concentrations considered in this study, each displayed in a different row. The orange-colored curves display contour curves of characteristic values of the damping rate, while the white curves correspond to the contour plot of $\text{Re}(P)=0$. The beta parameter is set at $\beta_e=0.01$ (left) and $\beta_e=0.10$ (right), assuming isothermy between species. }
    \label{fig:heat_k}
\end{figure*}

\begin{figure*}[ht]
    \centering
    \includegraphics[width=\textwidth]{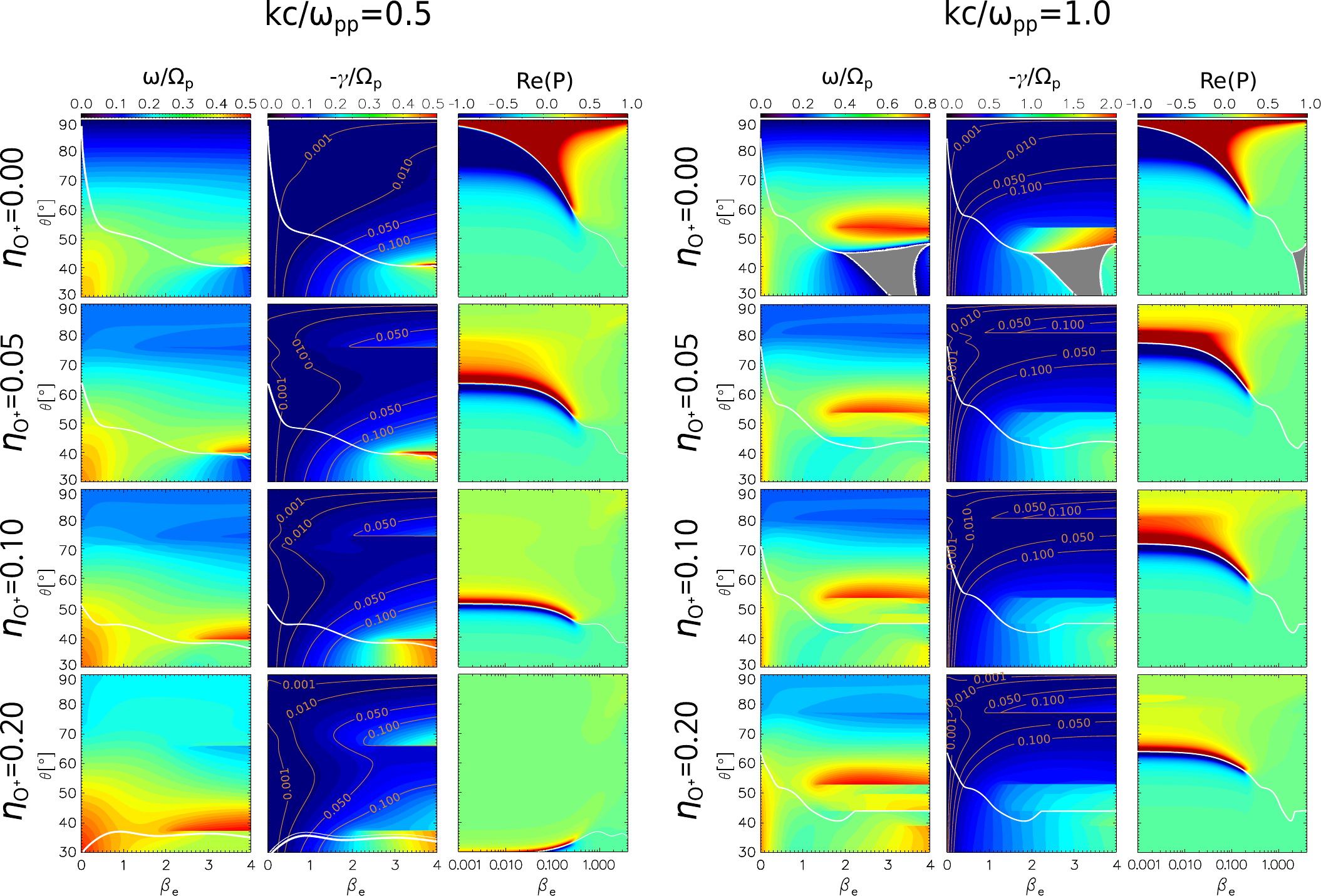}
    \caption{Heat maps of real frequency, negative damping rate, and real polarization in logarithmic scale as functions of the plasma beta for fixed wavenumbers of $k = 0.5$ (left) and $k = 1.0$ (right). As in Fig.  \ref{fig:heat_k}, the different rows correspond to the different oxygen ion concentrations, orange curves are contours of characteristic values of damping rate, and white curves are contours of $\text{Re}(P)=0$. The grey-shaded region indicates a domain of linear polarization.}
    \label{fig:heat_beta}
\end{figure*}

In order to extensively analyze the transition from negatively polarized EMIC waves to the KAW mode, we plot solutions (e.g., shown in Figs. \ref{fig:disp_angles} and \ref{fig:disp_angles_beta}) in the form of heat maps, such that we may consider the propagation at every possible wave-normal angle. Figure \ref{fig:heat_k} shows the Alfvénic mode properties for the frequency and polarization as functions of the wavenumber for plasma beta values of $\beta_e = 0.01$  and $\beta_e=0.10$. Figure \ref{fig:heat_beta}, displays the same quantities as functions of the beta parameter, with wavenumbers fixed at $ck/\omega_{pp}=0.5$ and $ck/\omega_{pp}=1.0$. In Fig. \ref{fig:heat_beta}, for all plots of the real part of the polarization (right column), the plasma beta appears in logarithmic scale, as in \citep{gary_1986}. This choice allows us to thoroughly analyze the transition from a cold to a warm plasma and facilitate insights into the behavior of the plasma's polarization for low beta values, which tend to be more common in the inner magnetosphere \citep{Cohen2017}.
Additionally, in Figs. \ref{fig:heat_k} and \ref{fig:heat_beta}, the white lines in all maps specify the contour curve of zero polarization, which indicates where the transition from EMIC (negative polarization) to KAW (positive polarization) occurs. We also plot the contours for characteristic values of the damping rate to identify the overlap between weakly damped waves and positive polarization at oblique angles (weakly damped KAWs solutions).

As previously noted in Figs. \ref{fig:disp_angles} and \ref{fig:disp_angles_beta} and further displayed in Figs. \ref{fig:heat_k} and \ref{fig:heat_beta}, we observe that the inclusion of oxygen ion populations tends to increase the damping rate of the Alfvénic waves, especially for lower wavenumbers in Fig. \ref{fig:heat_k} and near-perpendicular propagation in Fig. \ref{fig:heat_beta}. We also note that the contours of $\gamma/\Omega_p\leq -0.05$ are mostly unchanged when plotted as functions of the wave number in Fig.  \ref{fig:heat_k}. The same does not hold for the left panel of \ref{fig:heat_beta}, when the contours are plotted as a function of the beta parameter with $ck/\omega_{pp}=0.5$. This is something to be expected, as this value of the normalized wavenumber tends to be in the region of the damping rate bump caused by the inclusion of oxygen ions, as can be seen in Figs. \ref{fig:disp_angles} and \ref{fig:disp_angles_beta}. Nevertheless, we observe from both  Figs. \ref{fig:heat_k} and \ref{fig:heat_beta} that the inclusion of the heavier oxygen ions effectively increases the domain of weakly damped right-hand KAWs, as the $\text{Re}(P)=0$ contours change drastically with an increasing population of oxygen ions toward lower angles in all cases considered.

It is worth noting that the main results for the two plasma beta values considered in Fig. \ref{fig:heat_k} are very similar when it comes to the effect of the heavy ions on the dispersive properties of the waves. The main differences come from the solutions having lower frequencies, being less damped, and changing their polarization at greater angles for bigger wave numbers, which are all results that were to be expected. It is interesting to note, however, that when the plasma beta is small ($\beta_e=0.001$), the polarization of the waves tends to becomes linear rather than right-hand polarized as the waves transition from EMIC to KAW, as expected. For a plasma beta of $\beta_e=0.10$, this is no longer the case and the waves can become right-hand elliptically polarized, although linear polarization is maintained at near-perpendicular propagation.

When analyzing the dispersion properties as functions of the beta parameter, we see from Fig. \ref{fig:heat_beta} that the main results are again very similar. However, two main differences should be mentioned between the results considering the different fixed wave numbers. The first is that the lowering of the transition angle due to the presence of heavy ions is far more pronounced in the case where the wave number is smaller. This is to be expected when looking at Fig. \ref{fig:heat_k}, as the slope of the transition curve is much steeper when $ck/\omega_{pp}=0.5$ than it is for $ck/\omega_{pp}=1.0$. The second is the apparition of a wide domain where $\text{Re}(P)=0$ in the electron-proton case for  $ck/\omega_{pp}=1.0$. This region, shaded gray in the figure, is to be interpreted as a domain of linear polarization as the mode becomes aperiodic, which is also present in the other cases but for higher values of the plasma beta that are not pertinent to this study. A similar result was observed in Fig. 9 of \cite{Moya2015} for Alfvén wave branch asymptotic to $\Omega_{O^+}$, and it appears to be a feature from linear theory of propagation at oblique angles with high values of the plasma beta and wavenumber. Although this result and the effects that oxygen ions can play in the displacement of this linear polarization region are interesting, a detailed study and explanation of this phenomenon lie beyond the scope of this article and should be treated in future works. 

\begin{figure*}[ht]
    \centering
    \includegraphics[width=\textwidth]{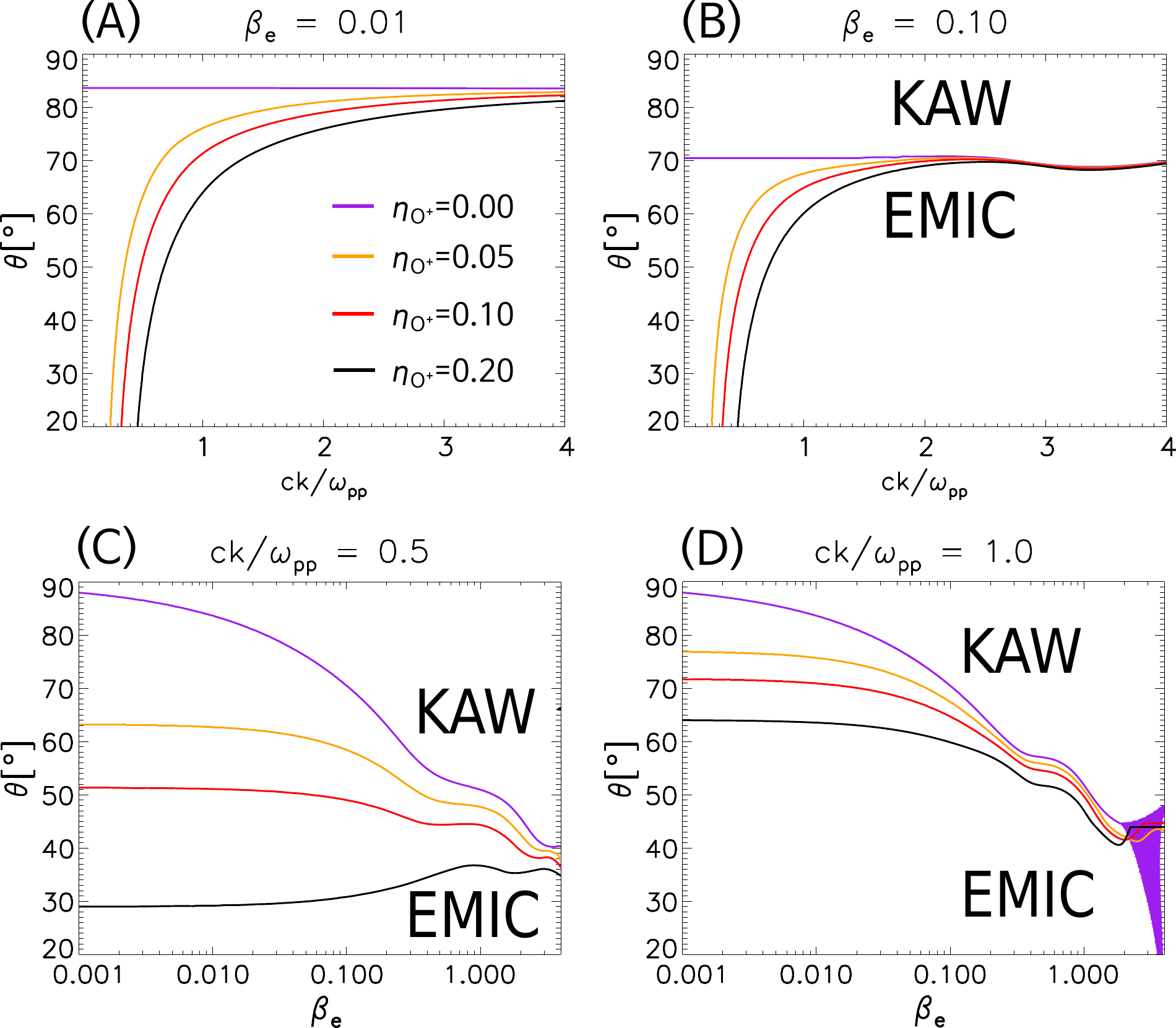}
    \caption{Contour plots of $\text{Re}(P)=0$ for Alfvén waves as a function of the wavenumber with \textbf{(A)} $\beta_e=0.01$ and \textbf{(B)} $\beta_e=0.10$; and as a function of the plasma beta parameter with \textbf{(C)} $ck/\omega_{pp}=0.5$ and \textbf{(D)} $ck/\omega_{pp}=1.0$. The solutions for the different oxygen ion concentrations considered are displayed in different colors. Above each curve, the solution is right-hand polarized (KAW) for a specific oxygen ion concentration, while below it is left-hand polarized (EMIC).}
    \label{fig:contour} 
\end{figure*}

Finally, from the results shown in Figs. \ref{fig:heat_k} and \ref{fig:heat_beta}, we extracted the contours for the $\text{Re}(P)=0$ level, displaying them in Fig. \ref{fig:contour} for the different oxygen ion concentrations, both as functions of the wavenumber (top panels) and plasma beta parameter (bottom panels). With these plots, we can better grasp the effect of the oxygen ion concentration on the transition from negatively to positively polarized Alfvénic waves, namely, the transition from EMIC waves to KAWs. The results show a clear tendency for a higher oxygen ion concentration to lower the transition angle from EMIC waves to KAW, which is more pronounced when the dependence on the beta parameter is analyzed for relatively small wavenumbers. The contours displayed in Figure \ref{fig:contour} provide strong evidence that, according to linear theory, the propagation of KAWs at wave-normal angles under 70º remains impossible in the magnetospheric environment for low beta values, such as the low-density case considered in \citet{Moya2015}, unless a significant population of oxygen ions is present in the plasma. As the propagation angle of KAW measured in the above article averages around 60º, propagation of right-hand polarized Alfvén waves must occur at even lower angles. Even for medium to high particle densities, this would be highly unlikely if the plasma consists only of electrons and protons. However, our results prove that this anomaly can be easily explained by considering oxygen ion populations of $\eta_{O^+}\sim 0.2$, which are consistent with those of the inner magnetosphere.


\section{Summary and conclusions}

We present an extensive study on the effect of oxygen ions on the dispersion properties of Aflvénic waves in a multi-species electron-proton-oxygen ion plasma with no temperature anisotropy. Considering inner magnetospheric-type parameters, we first demonstrated that an increasing oxygen ion population allows the propagation of KAW at remarkably low wave-normal angles by heavily displacing the transitional region from left-hand polarized EMIC waves to right-hand polarized KAW in phase-space toward lower wave-normal angles. These results are consistent with recently published case studies \citep{MOYA2021,Moya2022}, as well as with observations of KAW propagating at about 60º with respect to Earth's magnetic field in the inner magnetosphere \citep{Moya2015, Chaston2014}.

Secondly, by fixing the wavenumber and allowing the plasma beta parameter to vary, we show that the behavior of the EMIC-KAW transition curve in $\theta-\beta$ space drastically changes with the oxygen ion concentration. While an increase in the plasma beta facilitates the propagation of KAW in a plasma with oxygen ion concentrations of $\eta_{O^+}=0.00$, $0.05$, and $0.10$, for a higher oxygen ion concentration of $\eta_{O^+}=0.20,$ we see that an increase in the plasma beta parameter increases the transition angle from EMIC to KAW. However, for all the cases considered, the transition curve always lies below the case with less oxygen abundance for all values of the beta parameter. This decrease in the transition angle from EMIC to KAW is particularly pronounced for low-beta plasmas ($\beta\leq 1$), which tend to be fairly common configurations in  planetary magnetospheres\citep{thomsen_2010,Cohen2017}.

It is also worth noting that an increase in the oxygen ion concentration increases the damping rate of the waves, particularly for small wave numbers (Figs. \ref{fig:disp_angles}, \ref{fig:heat_k}). Thus, it is not surprising that Fig.  \ref{fig:heat_beta} shows a considerable decrease in the imaginary part of the frequency for $ck/\omega_{pp}=0.5$ for fixed beta values as the oxygen ion concentration increases. We emphasize, however, that (as shown in Figs. \ref{fig:heat_k}) the inclusion of significant concentrations of oxygen ions only modifies the damping rate of the waves by small amounts compared to the electron-proton case, with the $\gamma/\Omega_p\leq-0.05$ domain remaining practically untouched as $\eta_{O^+}$ increases. As seen in Fig.  \ref{fig:heat_beta}, the same is not valid when the dependence on the beta parameter is analyzed. Nevertheless, for $\bar{\beta}<2$, variations of the $\gamma/\Omega_p\geq-0.05$ domain are only slight between a $5\%$ and a $20\%$ oxygen ion concentration. This allows for considerable overlap between the regions of right-hand polarized and weakly damped Alfvénic waves for typical values of the beta parameter, allowing KAW to effectively propagate at smaller wave-normal angles than in the electron-proton case.

In summary, this article provides strong theoretical evidence for the existence of KAWs propagating at relatively low wave-normal angles in multi-species plasmas such as those present in regions of planetary magnetospheres of Earth, Saturn,  Jupiter and other planets. This behavior is consistent with satellite measurements of the inner terrestrial magnetosphere {reported} in the last decade, particularly with the Van Allen Probes \citep{Chaston2014,Moya2015}. We attribute these observations to the high abundance of oxygen ions in this region of space. Indeed, according to our results, without the presence of oxygen ions, there would not be KAWs to be observed at such angles.  Nonetheless, given that the kinetic dispersion relation of plasma waves depends not only on the wavenumber, concentration, and plasma beta, the effect of other parameters (such as temperature anisotropy and relative drift particle populations) are yet to be studied.
Furthermore, temperature anisotropies are known to give rise to kinetic micro-instabilities, which could compensate for the additional damping due to the presence of oxygen ions. The relative effect of ions with higher (such as He${}^+$, He${}^{2+}$, O${}^{2+}$, and highly ionized particles that are also present in the solar wind) or lower  (such as S${}^{+}$ or light ionized heavy metals) charge-to-mass ratios on the dispersive properties of KAW is also yet to be determined. Future studies should therefore extend and refine these results, particularly on the basis of more realistic conditions, and provide definitive explanations for observations of plasma phenomena  in the magnetosphere.

\begin{acknowledgements}
We thank the support of ANID, Chile, through National Doctoral Scholarship N$^\circ$ 21220616 (NVS), Fondecyt Grant 1191351 (PSM), and Fondecyt Initiation Grant 11201048 (RAL).
\end{acknowledgements}

%
%
\bibliographystyle{aa}
\bibliography{biblio_sorted}

\begin{appendix} 
\section{Details of the dispersion tensor}
Let us take a background velocity distribution function given by a bi-Maxwellian distribution with drift:

\begin{equation}
\label{EqA1}
    f_{0s}(v_\parallel,v_\perp)=\frac{n_{0s}\pi^{-3/2}}{\alpha_{\perp s}^2 \alpha_{\parallel s}}\exp \Bigg\{{-\frac{v_\perp^2}{\alpha_{\perp s}^2}-\frac{(v_\parallel-U_s)^2}{\alpha_{\parallel s}^2}}\Bigg\},
\end{equation}
where $n_{0s}$ is the mean number density of the $s$-th species, $\alpha_{\parallel s}=(2k_BT_{\parallel s}/m_s)^{1/2}$ and $\alpha_{\perp s}=(2k_BT_{\perp s}/m_s)^{1/2}$ are the thermal speeds of the species, and $T_{\parallel s}$ and $T_{\perp s}$ are the temperatures in the parallel and perpendicular direction with respect to the mean magnetic field $\mathbf{B}_0$. The drift velocity along the magnetic field is given by $U_s$, while $m_s$ and $k_B$ denote the particle mass of the $s$-th species and the Boltzmann constant, respectively. We note that the distribution function in (\ref{Eq1}) is a particular case presented here. The expressions utilized throughout this study can be easily obtained by taking $U_{s}=0$ and $\alpha_{\perp s}=\alpha_{\parallel s}=\alpha_s$ for all particle species.

The dispersion tensor of (\ref{disp}), obtained by integrating the first order perturbation term of the linearized Vlasov equation over velocity space in cylindrical coordinates, is given by \citep{Goldstein1985,Stix1992,Vinas_2000, Lopez_2017, Yoon2017, MOYA2021}:

\begin{equation}\label{Eq3}
    D=\begin{pmatrix}
    D_{xx}&iD_{xy}&D_{xz}\\
    -iD_{xy}&D_{yy}&iD_{yz}\\
    D_{xz}&-iD_{yz}&D_{zz}
    \end{pmatrix},
\end{equation}
with
\begin{equation}\label{Eq4}
      D_{xx}=1-\frac{c^2k_\parallel^2}{\omega^2}+\sum_s\frac{\omega_{ps}^2}{\omega^2}\sum_{\ell=-\infty}^{\infty}\frac{\ell^2\Lambda_\ell(\lambda_s)}{\lambda_s}\mathcal{A}_{\ell},
\end{equation}

\begin{equation}\label{Eq5}
      D_{xy}=\sum_s\frac{\omega_{ps}^2}{\omega^2}\sum_{\ell=-\infty}^{\infty} \ell \Lambda'_{\ell}(\lambda_s)\mathcal{A}_{\ell},
\end{equation}

\begin{equation}\label{Eq6}
    D_{xz}=\frac{c^2k_\perp k_\parallel}{\omega^2} -\sum_s  \frac{q_s}{|q_s|}\frac{\omega_{ps}^2}{\omega^2}\mu_s^{-1/2}\sum_{\ell=-\infty}^{\infty}\frac{\ell \Lambda_{\ell}(\lambda_s)}{\sqrt{2\lambda_s}}\mathcal{B}_{\ell},
\end{equation}\\

\begin{equation} \label{Eq7}
       D_{yy}=1-\frac{c^2k^2}{\omega^2} + \sum_s\frac{\omega_{ps}^2}{\omega^2} \sum_{\ell=-\infty}^{\infty} \Bigg[\frac{\ell^2\Lambda_{\ell}(\lambda_s)}{\lambda_s}-2\lambda_s\Lambda'_{\ell}(\lambda_s)\Bigg]\mathcal{A}_{\ell},
\end{equation}

\begin{equation} \label{Eq8}
      D_{yz}=\sum_{s}\frac{q_s}{|q_s|}\frac{\omega_{ps}^2}{\omega^2}\mu_s^{-1/2}\sum_{\ell=-\infty}^{\infty}\sqrt{\frac{\lambda_s}{2}}\Lambda'_{\ell}(\lambda_s)\mathcal{B}_{\ell},
\end{equation}
and
\begin{align} \label{Eq9}
      D_{zz}&=1-\frac{k_\perp^2c^2}{\omega^2}+2\sum_{s}\frac{\omega_{ps}^2}{\omega^2}\mu_s^{-1}\frac{U_s}{\alpha_{\parallel s}}\left[\frac{U_s}{\alpha_{\parallel s}}+2\xi_s\right]\nonumber\\
      &\hspace{3cm}+2\sum_{s}\frac{\omega_{ps}^2}{\omega^2}\mu_s^{-1}\sum_{\ell=-\infty}^{\infty}\Lambda_{\ell}(\lambda_s)\mathcal{C}_{\ell}.
\end{align}
 Here, we introduce the plasma frequency $\omega_{ps}=(4\pi n_{0s}q_s^2/m_s)^{1/2}$ and gyrofrequency $\Omega_s=q_s B_0/m_s c$ of the species, while $q_s$ is the charge of the $s$-th species of particles. Also, we define the functions:

 \begin{align}
\mathcal{A}_{\ell}&=(\mu-1)+\Big[\xi_{s}+(\mu-1)\zeta_{\ell s}\Big]\mathcal{Z}(\zeta_{\ell s}),\nonumber \\\mathcal{B}_{\ell}&=-2(\xi_{s} + \zeta_{\ell s}\mathcal{A}_{\ell})\quad \textnormal{ and }\quad  \mathcal{C}_{\ell}=\xi_s \zeta_{\ell s}+\left(\zeta_{\ell s}+\frac{U_s}{\alpha_{\parallel s}}\right)^2\mathcal{A}_{\ell}
 ,\end{align}
with the quantities

\begin{align} \label{Eq10}
   & \mu_s = \frac{\alpha_{\perp s}^2}{\alpha_{\parallel s}^2}=\frac{T_{\perp s}}{T_{\parallel s}},\quad \xi_{s} = \frac{\omega - k_{\parallel}U_s}{k_{\parallel}\alpha_{\parallel s}}, \quad \zeta_{\ell s}= \xi_{s}-\frac{\ell \Omega_s}{k_\parallel \alpha_{\parallel s}},\nonumber\\
    &\hspace{6.5cm}\quad\text{ and } \quad \lambda_s=\frac{k_{\perp}^2\alpha_{\perp s}^2}{2\Omega_s^2},
\end{align}
as well as the special functions

\begin{equation} \label{Eq11}
    \Lambda_\ell(x)=e^{-x}I_{\ell}(x) \qquad \text{  and  }\qquad Z(x)=\frac{1}{\sqrt{\pi}}\int_{-\infty}^{\infty}\frac{e^{-t^2}}{t-x}dt,
\end{equation}
where in (\ref{Eq11}) the $I_\ell$ is the integer index modified Bessel functions of the first kind, and $Z(x)$ corresponds to the plasma dispersion function.\\

Moreover, from (\ref{disp}), we can express the polarization of the electromagnetic waves defined in (\ref{polStix}) in terms of the components of the tensor in (\ref{Eq3}) as follows:
\begin{align}
   P(k)&=i\frac{E_{kx}}{E_{ky}} =\frac{D_{xy}D_{zz}+D_{xz}D_{yz}}{D_{xx}D_{zz}-D_{xz}^2}\text{sign}(\omega).\label{pol1}
\end{align}

\section{Identification of the Alfvén solution}

We identified the Alfvén solutions by comparing the obtained dispersion relations with those of MHD Afvén waves, given by $\omega_A=k_{\parallel}V_A$, where $V_A={B}/{\sqrt{4\pi n_{H^+}m_{H^+}}}$ is the Alfvén speed for H${}^+$ ions, and by considering the fact that these solutions are known to have $P(k)=iE_{kx}/E_{ky}>0$ at angles close to 90${}^\circ$. \\
Comparing our solutions to the MHD limit allows us to recognize the Alfvén solution from other modes in the same frequency range that satisfy the same condition for the scalar polarization in \ref{pol1}, such as the fast magnetosonic mode. The dispersion relation for this mode, in the classical MHD limit, is given by $\omega_{f}=ck\sqrt{\frac{V_s^2+v_A^2}{c^2+v_A^2}}$, where $V_s\approx \sqrt{\beta}V_A$ is the speed of sound in the plasma.   

\begin{figure}[ht]
    \centering
    \includegraphics[width=0.5\textwidth]{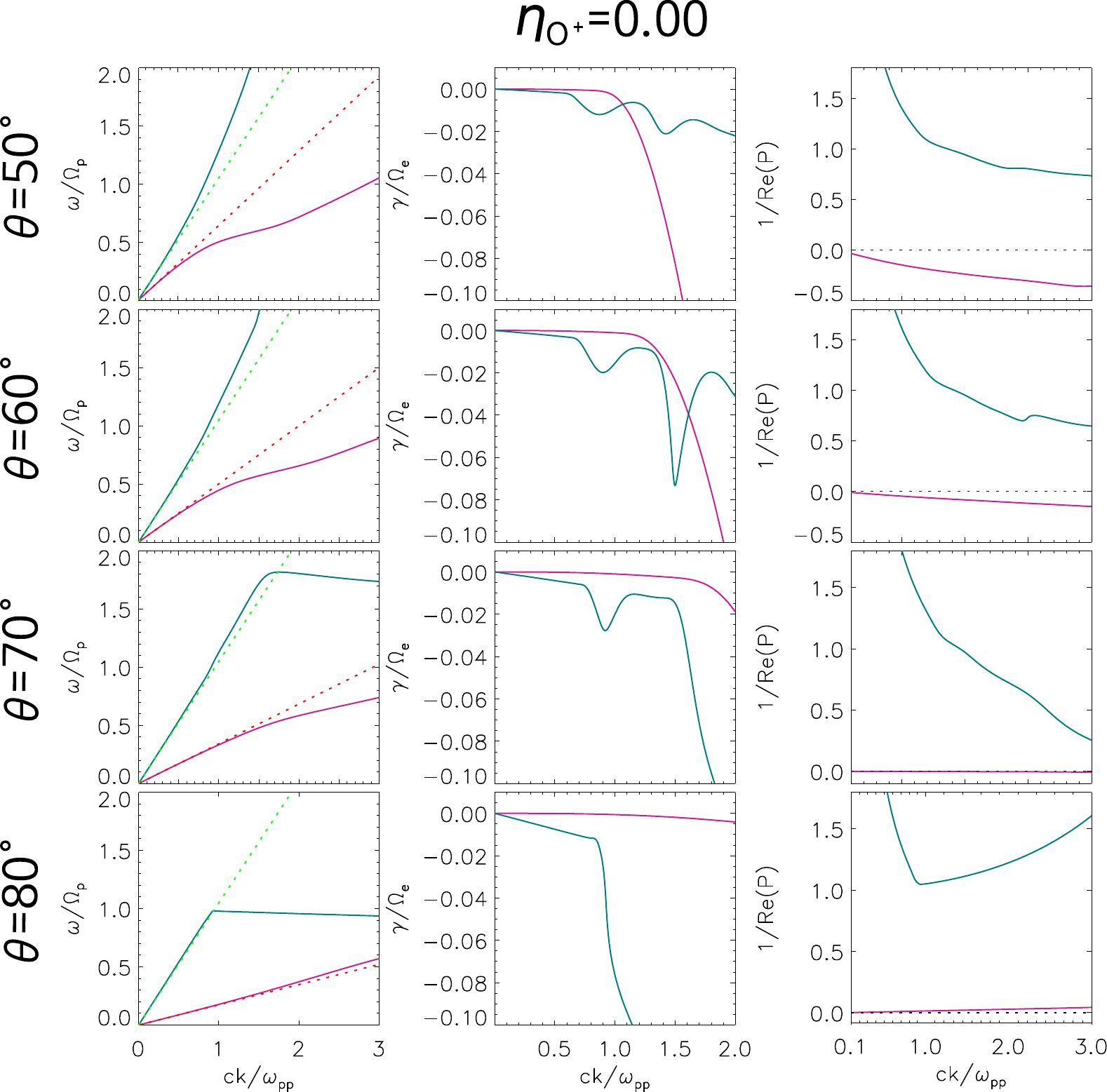}
    \caption{Dispersion relation of EMIC/KAW (continuous magenta) and fast magnetosonic-whistler waves (continuous teal) for different propagation angles in an electron-proton plasma. Dashed lines in the respective colors indicate the MHD dispersion relation.}
    \label{fig:B1}
\end{figure}

\begin{figure}[ht]
    \centering
    \includegraphics[width=0.5\textwidth]{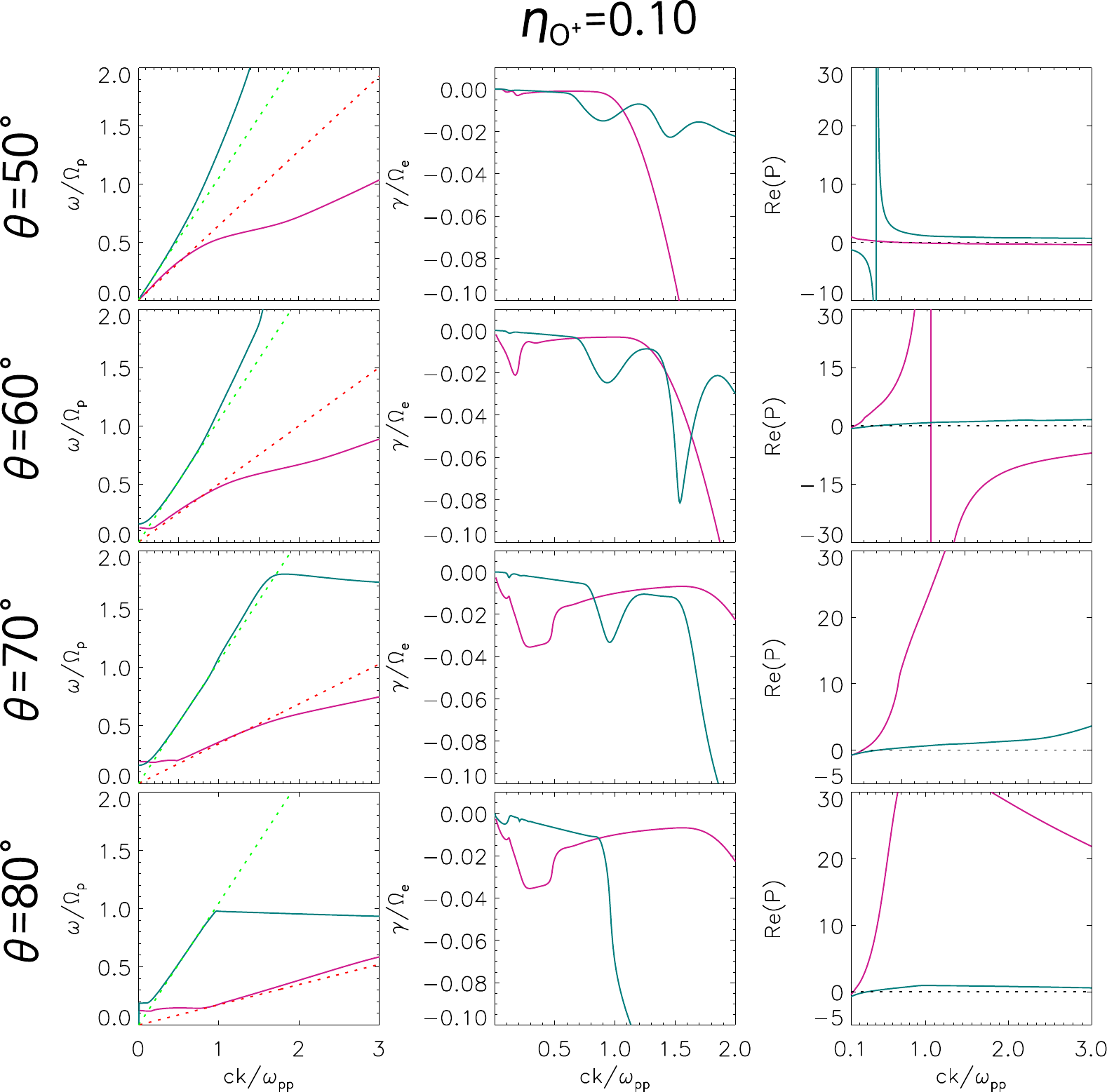}
    \caption{Dispersion relation of EMIC/KAW (continuous magenta) and fast magnetosonic-whistler waves (continuous teal) for different propagation angles in an electron-proton-O${}^+$ plasma with $\eta_{O^+}=0.10$. Dashed lines in the respective colors indicate the MHD dispersion relation.}
    \label{fig:B2}
\end{figure}

Figure \ref{fig:B1} shows the complex frequency and inverse real part of the polarization for Alfvén (continuous magenta) and fast magnetosonic (continuous teal) waves obtained from kinetic theory in an electron-proton plasma for propagation angles of 50$^\circ$, 60$^\circ$, 70$^\circ$, and 80$^\circ$. Dashed lines of similar colors indicate the MHD solutions mentioned above. Figure \ref{fig:B2} displays the same quantities in an electron-proton-O$^+$ plasma with $\eta_{O^+}=0.10$. The real part of the polarization is plotted instead of its inverse for convenience.

From Figure \ref{fig:B1}, we can see that the fast magnetosonic-whistler mode is completely decoupled from the Alfvén wave and, although both waves are right-hand polarized at higher propagation angles, the nature of their polarization is completely different; the polarization of the fast wave is closer to circular polarization, while the polarization of the KAW is nearly linear.

Figure \ref{fig:B2} shows that the real frequency of the waves is nearly unchanged by the inclusion of oxygen ions, except for the apparition of a small forbidden frequency range near $\omega=\Omega_{O^+}$, which causes the wave to split into two frequency bands. The imaginary frequency of the Alfvén mode is also only slightly changed by the apparition of a damping bump for small wavenumbers. We discuss both phenomena in depth in the analysis of Figure \ref{fig:disp_angles}. We also note that in this case as well, the solutions are decoupled. The polarization of the waves, however, changes its nature completely because of the effects of the heavy ion population, allowing the Alfvén waves to acquire right-hand polarization at lower propagation angles.

\section{Reproduction of transition curve by \citet{gary_1986}} 
\begin{figure}[h!]
    \centering
    \includegraphics[width=0.5\textwidth]{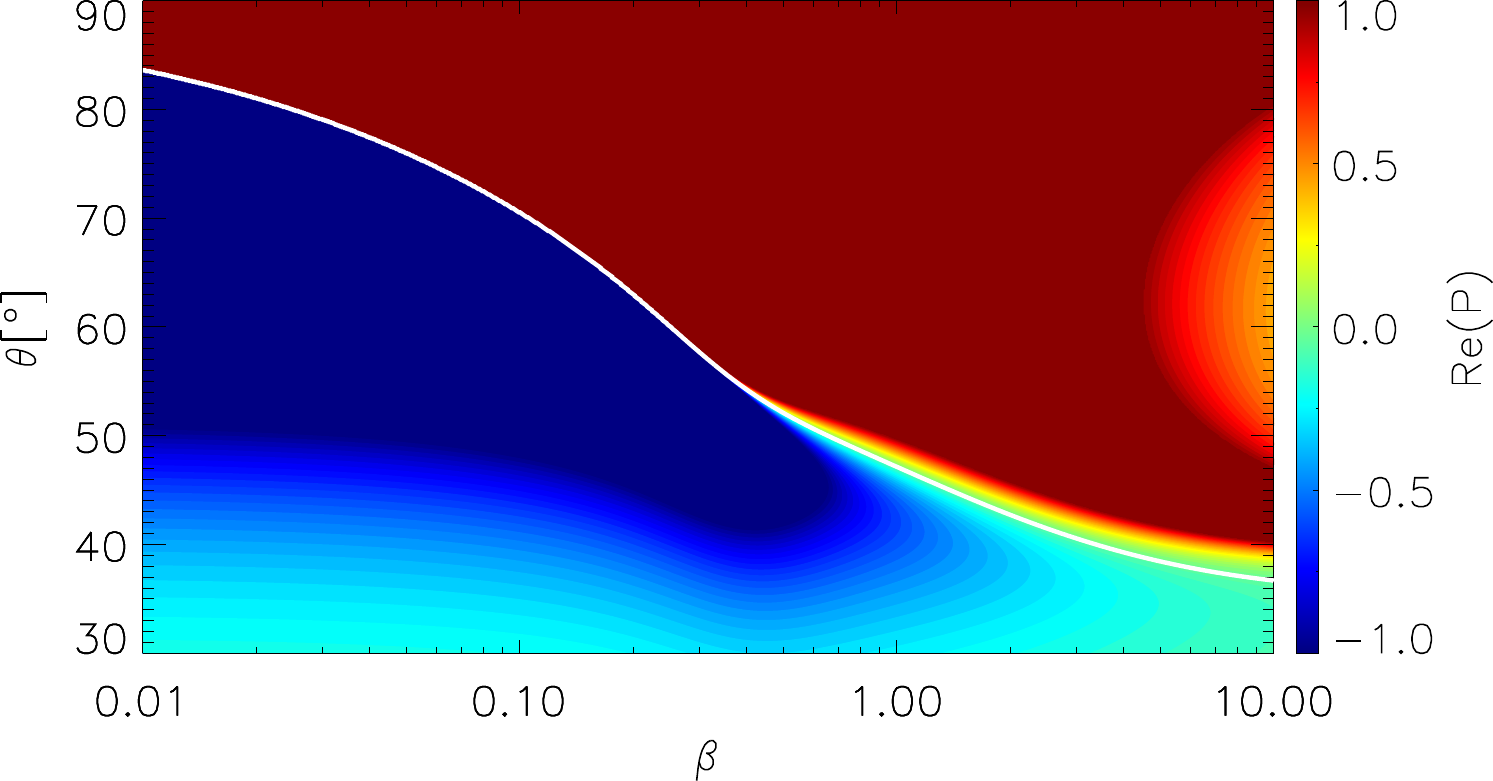}
    \caption{Heat map of the polarization of Alfvén waves in an electron-proton plasma for $ck/\omega_{pp}=0.05$.The white curve corresponds to the contour of $P=0$.}
    \label{fig:polk05}
\end{figure}

The methods utilized in this study allow us to replicate the curve of $\text{Re}(P)=0$ for obliquely propagating Alfvén waves displayed in Fig. 3 of \citet{gary_1986}. Figure \ref{fig:polk05} shows a heat map of the real part of the polarization of Alfvén waves in an electron-proton plasma for $ck/\omega_{pp}=0.05$ as a function of the wave-normal angle and plasma beta parameter. Since the plasma is considered isothermic, the plasma beta parameter is the same for both particle species, so the subscript is dropped. The white curve corresponds to the contour of $P=0$. As expected, this curve shows similar behavior to the purple curve in the bottom panel of Figure \ref{fig:contour}, since both contours are obtained from plasmas of the same composition, but different values of the wavenumber. As in the polarization maps in Fig. \ref{fig:heat_beta}, Fig. \ref{fig:polk05} shows how an isothermal increase in the plasma beta parameter can significantly reduce the transition angle from EMIC to KAW in an electron-proton plasma.

\end{appendix}

\end{document}